\documentclass[12pt]{article}  
\usepackage{epsfig}  
\def\beq{\begin{equation}}  
\def\eeq{\end{equation}}  
\def\eq{\beq\eeq}  
\def\beqn{\begin{eqnarray}}  
\def\eeqn{\end{eqnarray}}  
  
\relax

\def\l{\left}

\def\MS{\hbox{$\overline{\rm MS}$}}  
\def\pl#1#2#3{{\it Phys. Lett. } {\bf #1} (19#2) #3}  
\def\zp#1#2#3{{\it Z. Phys. } {\bf #1} (19#2) #3}

\def\pr#1#2#3{{\it Phys. Rev. } {\bf #1} (19#2) #3}  
\def\np#1#2#3{{\it Nucl. Phys. } {\bf #1} (19#2) #3}

\def    \hepph  #1 {{\tt hep-ph/#1}}  
\def    \hepex  #1 {{\tt hep-ex/#1}}  
\def\eq#1{eq.~(\ref{#1})}  
  
\newcommand{\sect}[1]{\setcounter{equation}{0}\section{#1}}  
  
\newcommand\sss{\scriptscriptstyle}  
  
\newcommand\as{\alpha_{\sss S}}

\newsavebox\tmpfig  
  
\begin{document}  
  
\pagestyle{empty}  
  
\begin{flushright}   
  
{\tt hep-ph/0006273}\\RM3-TH/00-09\\DFTT 27/00\\ GEF/TH-5-00\end{flushright}   
  
\begin{center}   
\vspace*{0.5cm}  
{\Large \bf Evolution of truncated moments \\   
of singlet parton distributions}\\  
\vspace*{1.5cm}   
{\bf Stefano Forte,$^{a,\,}$\footnote[1]{On leave from INFN, Sezione di  
Torino, Italy}  
Lorenzo Magnea,$^{b}$\\ Andrea Piccione$^{c,\;d}$} and   
{\bf Giovanni Ridolfi$^{d}$}\\  
\vspace{0.6cm}  
{}$^a$INFN, Sezione di Roma Tre\\  
Via della Vasca Navale 84, I-00146 Rome, Italy\\  
\vspace{0.4cm}  
{}$^b$Dipartimento di Fisica Teorica,  
Universit\`a di Torino and\\ INFN, Sezione di Torino \\  
Via P.~Giuria 1, I-10125 Torino, Italy\\  
\vspace{0.4cm}  
{}$^c$Dipartimento di Fisica, Universit\`a di Genova and\\ {}$^d$INFN,  
Sezione di Genova,\\  
Via Dodecaneso 33, I-16146 Genova, Italy\\  
\vspace*{1.5cm}  
  
{\bf Abstract}  
  
\end{center}  
  
\noindent  
We define truncated Mellin moments of parton distributions by  
restricting the integration range over the Bjorken variable to the  
experimentally accessible subset $x_0 \le x \le 1$ of the allowed  
kinematic range $0 \le x \le 1$.  We derive the evolution equations  
satisfied by truncated moments in the general (singlet) case in terms  
of an infinite triangular matrix of anomalous dimensions which couple  
each truncated moment to all higher moments with orders differing by  
integers.  We show that the evolution of any moment can be determined  
to arbitrarily good accuracy by truncating the system of coupled  
moments to a sufficiently large but finite size, and show how the  
equations can be solved in a way suitable for numerical  
applications. We discuss in detail the accuracy of the method in view  
of applications to precision phenomenology.  \\  
\vspace*{1cm}

\vfill  
\noindent  
  
\begin{flushleft} June 2000 \end{flushleft}   
\eject   
  
\setcounter{page}{1} \pagestyle{plain}  
  
\sect{The use of truncated moments}  
  
\noindent  
The needs of accurate phenomenology at current and future hadron  
colliders have recently led to the development of more refined tools  
for the QCD analysis of collider processes~\cite{qcd}.  A detailed  
understanding of parton distributions and their scaling violations is  
in particular an essential ingredient of such  
phenomenology~\cite{qcd,pdfrev}. Scaling violations of parton  
distributions are described by renormalization group equations for  
matrix elements of leading twist operators, whose anomalous dimensions  
are currently fully known to next-to-leading order~\cite{kis}, and to  
next-to-next-to-leading order for a handful of operators~\cite{verm}.  
Moments of parton distributions are related to moments of  
deep-inelastic structure functions by Wilson coefficients, which have  
also been computed up to next-to-next-to-leading order~\cite{willy}.  
Moments of structure functions, however, cannot be measured even  
indirectly, since they are defined as integrals over all kinematically  
allowed values of $x$, and thus require knowledge of the structure  
function for arbitrarily small $x$, {\it i.e.}  arbitrarily large  
energy.  
  
There is of course a well-known solution to this problem, which  
consists of using the Altarelli-Parisi equation~\cite{AP} to evolve  
parton distributions directly: the scale dependence of any parton  
distribution at $x_0$ is then determined by knowledge of parton  
distributions for all $x > x_0$, {\it i.e.}, parton evolution is  
causal.  In fact, through a judicious choice of factorization  
scheme~\cite{aem,cata} all parton distributions can be identified with  
physical observables, and it is then possible to use the  
Altarelli-Parisi equations to express the scaling violations of  
structure functions entirely in terms of physically observable  
quantities. It is, however, hard to measure local scaling violations  
of structure functions in all the relevant processes: in practice, a  
detailed comparison with the data requires the solution of the  
evolution equations  
  
What is usually done instead is to introduce a parametrization of  
parton distributions, and then solve the evolution equations in terms  
of this parametrization. The idea is that a parametrization fitted to  
the data will reproduce them in the experimentally accessible region,  
but it will also provide an extrapolation, so that the evolution  
equations can be solved easily, for instance taking Mellin  
moments. The results in the measured region should be independent of  
this extrapolation since, by the Altarelli-Parisi equation, measured  
scaling violations are independent of it. It has however become  
increasingly clear that in practice this procedure introduces a  
potentially large theoretical bias, whose exact size is very hard to  
assess~\cite{pdfrev,pdfer}. First, the very fact of adopting a  
specific functional form constrains not only the extrapolation in the  
unmeasured region, but also the allowed behavior at the boundary of  
the measured region. Especially when data are not very precise, it can  
be seen explicitly~\cite{abfr} that rather different results are  
obtained simply by changing the functional form used to parametrize  
parton distributions. Furthermore, it is very hard to assess the  
uncertainty on the best-fit functional form of the parton  
distributions, essentially because of the very nonlinear and indirect  
relation between the data and the quantity which is parametrized.  
Hence, the need to go through such a parametrization makes it very  
hard to assess the uncertainty on the desired result.  
  
Various methods to overcome these problem have been discussed in the   
literature. One possibility is to minimze the bias introduced by the
parton parametrization, by projecting parton distributions on an
optimized basis of 
functions, such as  suitable families of orthogonal   
polynomials~\cite{orto}. A more ambitious  proposal is to construct the  
probability functional for parton distributions through Bayesian  
inference applied on a Monte Carlo sampling of the relevant space of  
functions~\cite{pdfer}. The probability functional then summarizes in  
an unbiased way all the available experimental information, and can be  
used to determine the mean value and error on any physical  
observable. For many applications, however, there is a simpler,  
although more limited option, which consists of dealing directly with  
the experimentally accessible quantities. Indeed, consider two typical  
problems in the study of scaling violations: the determination of  
$\alpha_s$, and the determination of a moment of the gluon  
distribution, for instance the first moment of the polarized gluon  
distribution~\cite{abfr}, which gives the gluon spin  
fraction. Manifestly, in the latter case only the contribution to the  
moment from the measured region $x_0\le x\le 1$ is accessible  
experimentally, and it would be useful to be able to separate in a  
clean way the measured quantity from the extrapolation to the  
unmeasured region, which is necessarily based on assumptions. It is  
then natural to study the scaling violation of these measurable  
contributions to Mellin moments, {\it i.e.} truncated moments.  Similarly,  
$\alpha_s$ can be determined from the evolution equation of truncated  
moments~\cite{FM}, without any reference to the behavior of the  
structure function in the region in which it is not measured, and in  
principle without the need to resort to a specific functional  
parametrization.  
  
It turns out~\cite{FM} that scaling violations of truncated moments  
are described by a triangular matrix of anomalous dimensions which  
couple the $n$-th truncated moment to all truncated moments of order  
$n+k$, where $k$ runs over positive integers. This means that  
truncated moments share with full moments the property that evolution  
equations are ordinary first-order differential equations, and not  
integro-differential equations, as for the parton distributions  
themselves.  Unlike full moments, however, truncated moments can be  
measured without extrapolations. The price to pay for this is that  
their evolution equations do not decouple (unlike the case of full  
moments); however, the causal nature of the evolution implies that the  
evolution of each moment is only affected by higher order  
moments. Furthermore, the series of couplings to higher moments  
converges, and thus it can be truncated to any desired accuracy. The  
problem then reduces to the solution of a system of ordinary  
differential equations coupled by a triangular matrix, whose initial  
conditions are (measurable) truncated moments.  
  
This gives a simple solution to both problems mentioned above:  
$\alpha_s$ and the truncated moment of the gluon can be determined  
directly from the observed scaling violations, without having to go  
through an intermediate parton parametrization. A model independent  
error analysis can then be performed, provided only that data for the  
truncated moments and their errors are available. Of course, these  
could be extracted directly by summing over experimental bins, if a  
sufficiently abundant data set were available. In practice, however,  
it is more convenient to manipulate a smooth interpolation of the  
data. It turns out to be possible to do this without invoking an  
explicit functional parametrization for the measured structure  
functions, by constructing neural networks which are trained to  
simulate all available experimental information, including statistical  
and systematic errors and correlations. By means of these neural  
networks it is then easy to compute the observed truncated moments and  
their errors, which can be further used to perform a phenomenological  
analisys of scaling violations, free of theoretical bias. The  
application of the method of neural networks to the parametrization of  
structure functions is currently under way and will be presented in a  
separate publication.  
  
Evolution equations for truncated moments were presented in  
ref.~\cite{FM} in the simplest (nonsinglet) case, along with a  
preliminary study of the viability of the method. It is the purpose of  
this work to present a full treatment of the method of truncated  
moments, suitable for future phenomenological applications. In  
particular, in sect.~2 we will derive evolution equations for  
truncated moments in the general (singlet) case, and we will discuss  
their solution in a form which is suitable for numerical  
implementation. In sect.~3 we will consider numerical solutions of the  
evolution equations with typical quark and gluon distributions, and  
discuss the accuracy of the truncation of the infinite system of  
coupled evolution equations. Problems and techniques of interest for  
future phenomenological applications are discussed in sect.~4. The  
appendices collect various technical results which are needed for the  
actual implementation of the methods discussed here in an analysis  
code: in particular, we give explicit expressions for the NLO singlet  
splitting functions in the DIS scheme, and we list all the integrals  
needed to compute their truncated moments.

\sect{Evolution equations for truncated moments and their solutions}  
  
The $Q^2$ dependence of parton distributions $q(x, Q^2)$ is governed  
by the Altarelli-Parisi (AP) equations~\cite{AP}  
\beq  
\frac{d}{dt}~q(x, Q^2)=  
\frac{\as(Q^2)}{2 \pi} \int_x^1 \frac{d y}{y}   
P\left(\frac{x}{y}; \as(Q^2)\right) q(y, Q^2)~,  
\label{alpar}  
\eeq  
where $t = \log Q^2/\Lambda^2$.  In the non-singlet case, $q$ is  
simply one of the flavour non-singlet combinations of quark  
distributions, and $P$ the corresponding splitting function. In the  
singlet case, $q$ is a vector, whose components are the  
flavour-singlet combination of quark distributions,  
\beq  
\Sigma(x, Q^2) = \sum_{i=1}^{n_f} q_i(x, Q^2)  
\eeq  
and the gluon distribution $g(x, Q^2)$. Correspondingly, $P$ is a $2  
\times 2$ matrix of splitting functions, given as an expansion in  
powers of $\as$.  
  
As is well known, upon taking ordinary Mellin moments convolutions  
turn into ordinary products and evolution equations become ordinary  
first-order differential equations. By contrast, we are interested in  
the evolution of truncated moments, defined for a generic function  
$f(x)$ by  
\beq   
f_n(x_0) = \int_{x_0}^1 dx x^{n-1} f(x)\,.    
\eeq   
The corresponding evolution equations in the nonsinglet case were  
derived in ref.~\cite{FM}, which we will follow in presenting the  
generalization to the singlet case.  One finds immediately that the  
truncated moments of $q(x,Q^2)$ obey the equation  
\beq   
\frac{d}{d t} q_n(x_0, Q^2) = \frac{\as(Q^2)}{2\pi}   
\int_{x_0}^1 dy y^{n - 1} G_n  
\left(\frac{x_0}{y}\right) q(y, Q^2) \, ,  
\label{apsinglet}  
\eeq  
where  
\beq  
G_n(x) = \int_x^1 d z z^{n - 1} P(z)  
\label{kern}  
\eeq  
is perturbatively calculable as a power series in $\alpha_s$.  
  
Expanding $G_n(x_0/y)$ in powers of $y$ around $y=1$,  
\beq  
G_n \left(\frac{x_0}{y}\right) = \sum_{p=0}^\infty  
\frac{g_p^n (x_0)}{p!} (y - 1)^p\,;  
\;\;\;\;  
g_p^n (x_0)=\left[\frac{\partial^p}{\partial y^p}   
G_n  \left(\frac{x_0}{y}\right) \right]_{y = 1} \, ,  
\label{gntaylor}  
\eeq  
one obtains  
\beq  
\frac{d}{d t}q_n(x_0, Q^2)=  
\frac{\as(Q^2)}{2\pi}  
\sum_{p=0}^{\infty}\sum_{k=0}^p \frac{(-1)^{k+p}   
g_p^n (x_0)}{k! (p - k)!}\, q_{n + k}(x_0, Q^2)\,.  
\label{finsyst}  
\eeq  
The key step in the derivation of \eq{finsyst} is the term-by-term  
integration of the series expansion. This is allowed, despite the fact  
that the radius of convergence of the series in \eq{gntaylor} is $1 -  
x_0$, because the singularity of $G_n(x_0/y)$ at $y = x_0$ is  
integrable (this can be proven~\cite{Lebe} using the Lebesgue  
definition of the integral).  One can then express each power of  
$(y-1)$ using the binomial expansion, which leads to \eq{finsyst}.  
  
Equation~(\ref{finsyst}) expresses the fact that, while full moments of  
parton distributions evolve independently of each other, truncated  
moments obey a system of coupled evolution equations. In particular,  
the evolution of the $n^{th}$ moment is determined by all the moments  
$q_j$, with $j\ge n$.  In practice, the expansion in \eq{gntaylor},  
because of its convergence, can be truncated to a finite order $p =  
M$. The error associated with this procedure will be discussed in  
sect.~\ref{accuracy}. In this case, \eq{finsyst} can be rewritten as  
\beq  
\frac{d}{d t} q_n(x_0, Q^2) = \frac{\as(Q^2)}{2\pi}  
\sum_{k=0}^M c^{(M)}_{n k} (x_0) \, q_{n + k}(x_0, Q^2) \, ,  
\label{finsyst2}  
\eeq  
where  
\beq  
c^{(M)}_{nk}(x_0) = \sum_{p=k}^M \frac{(-1)^{p+k}   
g_{p}^n (x_0)}{k! (p-k)!}\,.  
\eeq  
To solve the system of equations (\ref{finsyst2}), it is necessary to  
include a decreasing number of terms ($M$, $M-1$, and so on) in the  
evolution equations for higher moments ($n+1$, $n+2$, \dots), obtaining  
$M+1$ equations for the $M+1$ truncated moments $\{q_n, \ldots, q_{n +  
M}\}$.  We will see in the next section that this approximation is  
fully justified.  In this case, the coupled system of evolution  
equations takes the form  
\beq  
\frac{d}{d\tau} \,q_k = \sum_{l=n}^{n+M} C_{kl}\, q_l \, ;   
\;\;\;\; n \leq k \leq n+M \, ,  
\label{finsyst3}  
\eeq  
with  
\beq  
\tau = \int_{t_0}^t dt' \, a(t') \, ;   
\;\;\;\; a(t) = \frac{\as(Q^2)}{2\pi}\, ,  
\label{tau}  
\eeq  
where  $C$ is now a triangular matrix:  
\beq  
\left\{  
\begin{array}{cccc}  
C_{k l} & = & c_{k,l - k}^{(M - k + n)} & (l \geq k)~~, \\  
C_{k l} & = & 0 &(l < k)~~.  
\end{array}  
\right.  
\label{matr0}  
\eeq  
  
In the nonsinglet case, discussed in ref.~\cite{FM}, the matrix   
elements $C_{k l}$ are just numbers, and the matrix $C$ in  
eq.~(\ref{matr0}) is triangular, which makes it easy to solve  
\eq{finsyst3} perturbatively.  In the singlet case, all the steps  
leading to \eq{matr0} are formally the same, but now each entry  
$C_{kl}$ is a $2 \times 2$ matrix. As a consequence, the matrix $C$,  
which is given in terms of partial moments of the evolution kernels,  
is no longer triangular, but has nonvanishing $2 \times 2$ blocks  
along the diagonal.  This problem can be circumvented, by writing the  
perturbative expansion of $C$ as  
\beq  
C = C_0 + a C_1 + \ldots \, = (A_0 + B_0) + a ( A_1 + B_1) + \ldots,  
\eeq  
where $A = A_0 + a A_1$ is block-diagonal, with $2 \times 2$ blocks on its   
diagonal,  
\beq  
A_{kl} = C_{kk} \delta_{kl} \, ,  
\eeq  
while $B = B_0 + a B_1$, considered as a matrix of $2\times 2$ blocks,  
is upper-triangular with vanishing diagonal entries. Now one can  
define a matrix $S$ that diagonalizes $A_0$,  
\beq  
S A_0 S^{-1} = \mbox{diag} (\gamma_1, \ldots, \gamma_{2M}) \, .  
\eeq  
Clearly, $S$ is $\tau$-independent, block-diagonal, and easily  
computed. Equation~(\ref{finsyst3}) can then be rewritten as  
\beq  
\frac{d}{d\tau}\,{\tilde q} = T \,{\tilde q}\,,  
\label{finsyst4}  
\eeq  
where ${\tilde q} = S \, q$ and $T = S C S^{-1}$.    
  
The new evolution matrix $T$ is triangular at leading order (with the  
same eigenvalues as $A_0$).  This is enough to solve the evolution  
equation to next-to-leading order, as in the nonsinglet case of  
ref.~\cite{FM}. The general solution is worked out in detail in  
Appendix A; the result is  
\beq  
{\tilde q}(\tau) = U(T, \tau) \, {\tilde q} (0) \, ,  
\label{qevol}  
\eeq  
where  
\beqn  
\label{singletsoltext}  
U_{ij}(T, \tau) & =  & R_{ik}^{-1} \left\{ \delta_{kl}  
\left(\frac{a(0)}{a(\tau)}\right)^{\gamma_l/b_0} \right. \\  
& + & \left. \frac{ {\hat T_1}^{kl} - b_1 \gamma_l \delta_{kl}}  
{\gamma_k - \gamma_l + b_0} \left[  
a(0) \left( \frac{a(0)}{a(\tau)} \right)^{\gamma_k/b_0}  
- a(\tau) \left( \frac{a(0)}{a(\tau)} \right)^{\gamma_l/b_0}   
\right] \right\} R_{lj} \, . \nonumber  
\eeqn  
In eqs.~(\ref{qevol},\ref{singletsoltext}), $T$ is expanded as  
$T=T_0+a T_1$; $R$ is the matrix which diagonalizes $T_0$, $R T_0  
R^{-1}=\mbox{diag}(\gamma_1,\dots\gamma_{2M})$; finally, $\hat T_1 = R T_1  
R^{-1}$.  
  
The matrix $R$ can be computed recursively, using the technique applied  
in ref.~\cite{FM} and proven in Appendix~B.  One finds  
\beqn  
& & R_{i j} = \frac{1}{\gamma_i - \gamma_j}  
              \sum_{p = i}^{j - 1} R_{i p} \, T_0^{p j}~,  
\label{trir}\\  
& & R^{-1}_{i j}=\frac{1}{\gamma_j - \gamma_i}  
                 \sum_{p = i + 1}^j T_0^{i p}\,R^{-1}_{p j},  
\label{invtrir}  
\eeqn  
which, together with the conditions $R_{ii}=1$ and $R_{ij}=0$ when  
$i>j$, determine the matrix $R$ completely.  
  
The general solution for the parton distributions is then  
\beq  
\label{generalsolution}  
q(\tau) = U(C, \tau) \, q(0) \, ,  
\eeq  
where  
\beq  
U(C, \tau) = S^{-1} U(T,\tau) S \, .  
\eeq  
The splittting functions and partial moment integrals which should be  
used in eq.~(\ref{singletsoltext}) in order compute this solution  
explicitly are listed in Appendices~C and D.  
  
For the sake of completeness, we describe a different method to solve  
\eq{finsyst2}. It is immediate to check that the matrix  
\beq  
\label{U}  
U (C, \tau) = I + \sum_{n = 1}^\infty \int_0^\tau d\tau_1 \ldots   
\int_0^{\tau_{n - 1}} d \tau_n C(\tau_1) \ldots C(\tau_n)  
\eeq  
obeys the differential equation  
\beq  
\frac{d}{d \tau} \, U(C, \tau) = C U(C, \tau)~,  
\eeq  
with the initial condition $U(C, 0) = I$. In general, \eq{U} is not  
very useful, since it involves an infinite sum.  In the present case,  
however, the infinite sum collapses to a finite sum.  To see this,  
consider again the decomposition $C = A + B$, where $A$ is  
block-diagonal and $B$ is upper-triangular.  It is easy to prove that  
\beq  
\label{newsol}  
U(C, \tau) = U(A, \tau) U({\tilde B}, \tau) \, ,  
\eeq  
where  
\beq  
{\tilde B} = U^{-1}(A, \tau) B U(A, \tau) \, .  
\label{tilb}  
\eeq  
Since $A$ is block diagonal, $U(A,\tau)$ is also block-diagonal, and  
it can be computed perturbatively using the procedure described in  
Appendix A. Furthermore, once $U(A)$ is known, the upper-triangular  
matrix $\tilde B$ can be computed through eq.~(\ref{tilb}). Now one  
can use the fact that upper-triangular matrices have the property that  
their $M$-th power vanishes. Hence, the solution can be expressed as  
the finite sum  
\beq  
U({\tilde B},\tau) = I + \sum_{n = 1}^{M - 1} \int_0^\tau d \tau_1   
\ldots \int_0^{\tau_{n - 1}} d \tau_n  
{\tilde B}(\tau_1) \ldots {\tilde B}(\tau_n) \, ,  
\eeq  
and from the knowledge of $U(\tilde B)$ and $U(A)$ one can determine  
the solution to the evolution equations explicitly.

\sect{Numerical methods and their accuracy}  
\label{accuracy}  
\def\R{{\cal R}}  
  
In this section we will assess the accuracy of our method when the  
series of contributions to the right-hand side of the evolution  
equation (\ref{finsyst}) is approximated by retaining a finite number  
$M$ of terms. The loss of accuracy due to this truncation is the price  
to pay for eliminating the dependence on parton parametrizations and  
extrapolations in the unmeasured region. However, unlike the latter  
uncertainties, which are difficult to estimate, the truncation  
uncertainty can be simply assessed by studying the convergence of the  
series.  A reasonable goal, suitable for state-of-the-art  
phenomenology, is to reproduce the evolution equations to about $5\%$  
accuracy: indeed, we expect the uncertainties related to the  
parametrization of parton distributions in the conventional approach  
to be somewhat larger ($\sim 10\%$)\footnote{Notice that this is {\it  
not} the uncertainty associated with evolution of a {\it given}  
parametrization with, say, an $x$-space code; rather, it is the  
uncertainty associated with the {\it choice} of the parametrization,  
and with the bias it introduces in the shape of the  
distribution.}. Notice that there is no obstacle to achieve a higher  
level of precision when necessary, by simply including more terms in  
the relevant expansions.  To this level of accuracy it is enough to  
study the behavior of the leading order contribution to the evolution  
equation: indeed, next-to-leading corrections to the anomalous  
dimension are themselves of order $10\%$. We have verified explicitly  
that the inclusion of the next-to-leading corrections does not affect  
our conclusions.  
  
We can compare the exact evolution equation~(\ref{finsyst}) with its  
approximate form, eq.~(\ref{finsyst2}), by defining the percentage error  
on the right-hand side of the evolution equations for the quark  
nonsinglet, singlet and gluon:  
\beqn  
&&\R_{n,M}^{\sss NS} = \frac{1}{{\cal N}_{\sss NS}}  
\int_{x_0}^1 dy~y^{n-1} \l[G^{\sss NS}_n \l(\frac{x_0}{y}\right)   
- \sum_{k=0}^M y^k  c^{\sss NS}_{nk} \right] q^{\sss NS}(y,Q^2)\,,\\  
&&\R_{n,M}^\Sigma = \frac{1}{{\cal N}_\Sigma}  
\int_{x_0}^1 dy~y^{n-1} \l\{\l[G^{qq}_n \l(\frac{x_0}{y}\right)   
- \sum_{k=0}^M y^k  c^{qq}_{nk} \right] \Sigma(y,Q^2) \right.   
\nonumber\\  
&&\phantom{aaaaaaaaaaaaaaaaaaa}  
+\l.\l[G^{qg}_n\l(\frac{x_0}{y}\right) - \sum_{k=0}^M y^k  c^{qg}_{nk}\right]   
g(y,Q^2)\right\},  
\\  
&&\R_{n,M}^g = \frac{1}{{\cal N}_g}  
\int_{x_0}^1 dy~y^{n-1} \l\{\l[G^{gq}_n \l(\frac{x_0}{y}\right)   
- \sum_{k=0}^M y^k  c^{gq}_{nk} \right] \Sigma(y,Q^2) \right.   
\nonumber\\  
&&\phantom{aaaaaaaaaaaaaaaaaaa}  
+\l. \l[G^{gg}_n\l(\frac{x_0}{y}\right) - \sum_{k=0}^M y^k  c^{gg}_{nk}\right]   
g(y,Q^2)\right\},  
\eeqn  
where ${\cal N}_{{\sss NS},\Sigma,g}$ are the exact right-hand sides  
of the evolution equation~(\ref{finsyst}).  We study the dependence of  
the percentage error on the value of $M$ for typical values of the  
cutoff $x_0$ and for representative choices of test parton  
distributions. In particular, we parametrize parton distributions as  
\beq  
q(x,Q^2) = N x^{-\alpha} (1-x)^{\beta} \, .  
\label{distr}  
\eeq  
We begin by choosing, as a representative case, $\beta=4$ and  
$\alpha=1$ for the singlet distributions and $\alpha=0$ for the  
nonsinglet. The nonsinglet is assumed to behave qualitatively as  
$q_{NS}\sim x g\sim x\Sigma$, in accordance with the behavior of the  
respective splitting functions. Furthermore, the normalization factors  
$N$ for the singlet and gluon are fixed by requiring that the second  
moments of $\Sigma(x,Q^2)$ and $g(x,Q^2)$ are in the ratio $0.6/0.4$,  
which is the approximate relative size of the quark and gluon momentum  
fractions at a scale of a few GeV$^2$.  We will then show that  
changing the values of $\alpha$ and $\beta$ within a physically  
reasonable range does not affect the qualitative features of our  
results.  
  
The accuracy of the truncation of the evolution equation is determined  
by the convergence of the expansion in \eq{gntaylor}. Because this  
expansion is centered at $y=1$, and diverges at $y=x_0$, the small $y$  
region of the integration range in \eq{apsinglet} is poorly reproduced  
by the expansion.  Hence, even though the series in \eq{finsyst}  
converges, as discussed in sect.~2, the convergence will be slower for  
low moments, which receive a larger contribution from the region of  
integration $y\sim x_0$. In fact, for low enough values of $n$, the  
convolution integral on the right-hand side of the evolution equation  
(\ref{apsinglet}) does not exist: this happens for the same value for  
which the full moment of the structure function does not exist, {\it  
i.e.} $n\le1$ in the unpolarized singlet and $n\le0$ in the  
unpolarized nonsinglet and in the polarized case.  Therefore, we  
concentrate on the lowest existing integer moments of unpolarized  
distributions, {\it i.e.}  the cases $n=2,3$ for the singlet  
distributions, and correspondingly $n=1,2$ for the nonsinglet, which  
are the cases in which the accuracy of the truncation will be worse.  
  
The values of $\R_{n,M}^{{\sss NS},\Sigma,g}$, computed at leading  
order with $x_0=0.1$, are shown in table~\ref{rhs01}.  
\begin{table}[htm]  
\begin{center}  
\begin{tabular}{|r||r|r|r||r|r|r|} \hline  
\multicolumn{7}{|c|}{$x_0=0.1$}\\ \hline  
M&  $\R_{1,M}^{\sss NS}$ & $\R_{2,M}^\Sigma$ & $\R_{2,M}^g$   
& $\R_{2,M}^{\sss NS}$ & $\R_{3,M}^\Sigma$ & $\R_{3,M}^g$ \\  
\hline  
\hline  
 5  & 0.63 &  0.43  & 0.55  &  0.16  &  0.12  &  0.16   \\ \hline      
10  & 0.49 &  0.36  & 0.38  &  0.13  &  0.10  &  0.12   \\ \hline  
20  & 0.34 &  0.27  & 0.26  &  0.10  &  0.08  &  0.08   \\ \hline      
40  & 0.20 &  0.17  & 0.17  &  0.06  &  0.05  &  0.05   \\ \hline             
70  & 0.12 &  0.10  & 0.10  &  0.04  &  0.03  &  0.03   \\ \hline       
100 & 0.09 &  0.07  & 0.07  &  0.03  &  0.02  &  0.02   \\ \hline       
150 & 0.06 &  0.05  & 0.05  &  0.02  &  0.01  &  0.01   \\ \hline       
\end{tabular}  
\end{center}  
\caption  
{\it Values of $\R_{n,M}^{{\sss NS},\Sigma,g}$ for $x_0=0.1$  
and different values of $n$ and $M$.}  
\label{rhs01}  
\end{table}  
The table shows that nonsinglet moments of order $n$ behave as singlet  
moments of order $n - 1$. This is a consequence of the fact that, as  
discussed above, the convergence of the expansion is determined by the  
singularity of the integrand $G_n\left({x_0/y}\right) q(y) $ of  
\eq{apsinglet} as $y\to x_0$; near $y = x_0$, the function  
$G_n\left({x_0/y}\right)$ is well approximated by the singular  
contribution $\log\left(1-{x_0/y}\right)$, while parton distributions  
carry an extra power of $y^{-1}$ in the singlet case in comparison to  
the nonsinglet.  We also observe in table~1 that, as expected, the  
convergence is slower for the lowest moments, and rapidly improves as  
the order of the moment increases.  This rapid improvement is a  
consequence of the fact that the convergence of the expansion of  
$G(x_0/y)$ is only slow in the immediate vicinity of the point  
$y=x_0$, and the contribution of this region to the $n$-th moment is  
suppressed by a factor of $x_0^{n-1}$.  Due to this fast improvement,  
the approximation introduced by including one less term in the  
expansion as the order of the moment is increased by one, which is  
necessary to obtain the closed system of evolution equations  
(\ref{finsyst3}), is certainly justified.  
  
The 5\% accuracy goal which we set to ourselves requires the inclusion  
of more than 100 terms for the lowest moment, but only about 40 terms  
for the next-to-lowest. The computation of series with such a large  
numer of contributions does not present any problem, since the  
splitting functions are known and their truncated moments are easily  
determined numerically. The implications of this requirement for  
phenomenology will be discussed in the next section.  
  
\begin{figure}[htbp]
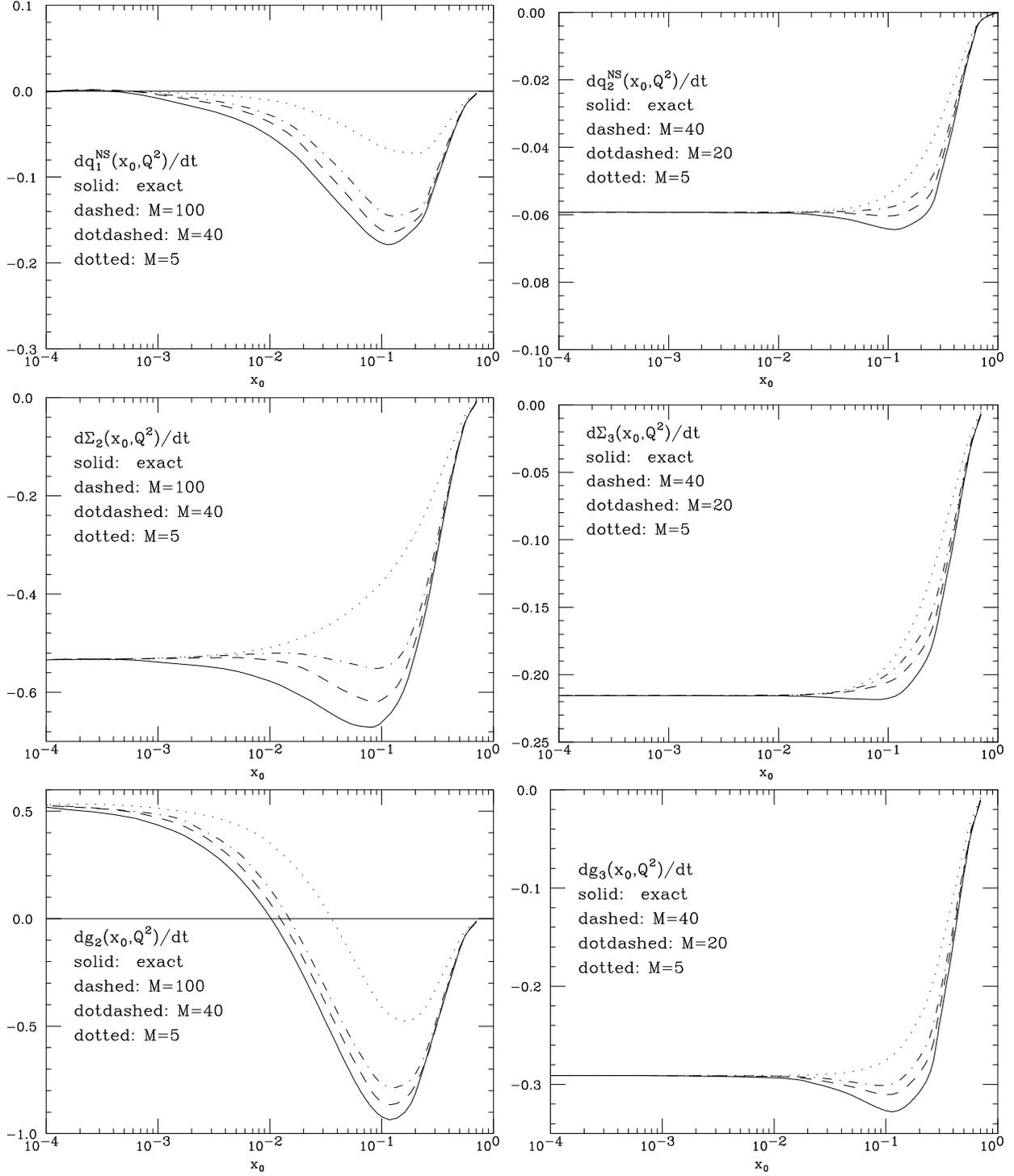
  
\begin{center}  
\epsfig{figure=derns1vsx0.ps,width=0.48\textwidth}  
\epsfig{figure=derns2vsx0.ps,width=0.48\textwidth}  
\epsfig{figure=derq2vsx0.ps,width=0.48\textwidth}  
\epsfig{figure=derq3vsx0.ps,width=0.48\textwidth}  
\epsfig{figure=derg2vsx0.ps,width=0.48\textwidth}  
\epsfig{figure=derg3vsx0.ps,width=0.48\textwidth}  
\end{center}  
\caption{{\it Right-hand sides of the evolution equations for the  
first and second truncated moments of the nonsinglet distribution, and  
for the second and third moments of singlet distributions. The overall  
scale is set by $\alpha_s(2\hbox{GeV}^2)$.}}  
\label{fig:RHS}  
\end{figure}

We can now study the dependence of these results on the value of the  
truncation point $x_0$ by plotting the exact and approximate  
right-hand side of the evolution equations as a function of $x_0$,  
as shown in fig.~\ref{fig:RHS}.  
The figures show that the case $x_0 = 0.1$ studied in table~1 is a  
generic one between the two limiting (and physically uninteresting)  
cases $x_0 = 0$ and $x_0 = 1$, where the approximation is exact.  In  
fact, with this particular choice of parton distributions, $x_0 = 0.1$  
is essentially a worst case and the error estimates of table~1 are  
therefore conservative.  
  
An interesting feature of these plots is the presence of zeroes of the  
lowest moment evolution at $x_0 = 0$ in the nonsinglet and around  
$x_0 \approx 10^{-2} $ in the gluon case. The physical origin of these  
zeroes is clear. At leading order, the first nonsinglet full moment  
does not evolve.  On the other hand, the second gluon full moment  
grows with $Q^2$, while higher gluon full moments decrease, {\it i.e.}  
the gluon distribution decreases at large $x$; this implies that the  
second truncated moment of the gluon must decrease for a high enough  
value of the cutoff $x_0$, while it must increase for very small  
$x_0$; its derivative is thus bound to vanish at some intermediate  
point. Of course, the phenomenology of scaling violations (such as a  
determination of $\alpha_s$) cannot be performed at or close to these  
zeroes, where there is no evolution. From the point of view of a  
truncated moment analysis, this means that the value of $x_0$ should  
be chosen with care in order to avoid these regions.  
  
Finally, in table~\ref{confev01} we study the dependence of our  
results on the form of the parton distributions, by varying the  
parameters $\alpha$ and $\beta$ within a reasonable range.  
\begin{table}[htm]  
\begin{center}  
\begin{tabular}{|r|r||r|r||r|r|} \hline  
\multicolumn{6}{|c|}{$n=2$, $x_0=0.1$}\\ \hline  
$\alpha$ & $\beta$ &$\R_{n,20}^\Sigma$   
              &$\R_{n,70}^\Sigma$ &$\R_{n,20}^g$ &$\R_{n,70}^g$ \\  
\hline  
\hline  
1.5 & 2.0 & 0.26 & 0.10 & 0.27 &  0.11\\  
\hline  
1.0 & 2.0 & 0.20 & 0.07 & 0.23 &  0.09\\  
\hline  
0.5 & 2.0 & 0.14 & 0.05 & 0.18 & 0.07\\  
\hline  
1.5 & 4.0 & 0.32 & 0.12 & 0.30 &  0.12\\  
\hline  
1.0 & 4.0 & 0.27 & 0.10 & 0.26 &  0.10\\  
\hline  
0.5 & 4.0 & 0.22 & 0.08 & 0.22 &  0.09\\  
\hline  
1.5 & 6.0 & 0.36 & 0.14 & 0.32 &  0.14\\  
\hline  
1.0 & 6.0 & 0.32 & 0.12 & 0.29 &  0.12\\  
\hline  
0.5 & 6.0 & 0.27 & 0.10 & 0.26 &  0.10\\  
\hline  
\hline  
\multicolumn{6}{|c|}{$n=3$, $x_0=0.1$}\\ \hline  
1.5 & 2.0 & 0.07 & 0.03 & 0.07 &  0.03\\  
\hline  
1.0 & 2.0 & 0.04 & 0.02 & 0.04 &  0.02\\  
\hline  
0.5 & 2.0 & 0.02 & 0.01 & 0.02 &  0.01\\  
\hline  
1.5 & 4.0 & 0.12 & 0.05 & 0.11 &  0.05\\  
\hline  
1.0 & 4.0 & 0.08 & 0.03 & 0.08 &  0.03\\  
\hline  
0.5 & 4.0 & 0.05 & 0.02 & 0.05 &  0.02\\  
\hline  
1.5 & 6.0 & 0.16 & 0.07 & 0.15 &  0.06\\  
\hline  
1.0 & 6.0 & 0.12 & 0.05 & 0.12 &  0.05\\  
\hline  
0.5 & 6.0 & 0.09 & 0.03 & 0.08 & 0.03  \\  
\hline  
\end{tabular}  
\end{center}  
\caption  
{\it Values of $\R_{n,M}^\Sigma$ and $\R_{n,M}^g$ for $x_0=0.1$  
and different choices of parton distribution parameters.}  
\label{confev01}  
\end{table}  
Of course, parton distributions which are more concentrated at small  
$y$ give rise to slower convergence. However, we can safely conclude  
that the effect of varying the shape of parton distributions is  
generally rather small. We have also verified that varying the  
relative normalization of the quark and gluon distributions has a  
negligible effect on the convergence of the series, even though it may  
change by a moderate amount the position of the zeroes in gluon  
evolution discussed above.  
  
\sect{Techniques for phenomenological applications}  
\label{pheno}  
  
So far we have discussed scaling violations of parton distributions.  
In a generic factorization scheme, the measured structure functions  
are convolutions of parton distributions and coefficient functions.  
When taking moments, convolutions turn into ordinary products and  
moments of coefficient functions are identified with Wilson  
coefficients. In the present case, however, as shown in  
eqs.~(\ref{apsinglet}-\ref{finsyst}), truncated moments turn  
convolutions into products of triangular matrices. Hence, in a generic  
factorization scheme, truncated moments of parton distributions are  
related to truncated moments of structure functions by a further  
triangular matrix of truncated moments of coefficient functions.  
  
This complication can be avoided by working in a parton  
scheme~\cite{aem}, where the quark distribution is identified with the  
structure function $F_2$. This still does not fix the factorization  
scheme completely in the gluon sector. One way of fixing it is to use  
a ``physical'' scheme, where all parton distributions are identified  
with physical observables~\cite{cata}. This may eventually prove the  
most convenient choice for the sake of precision phenomenology, once  
accurate data on all the relevant physical observables are  
available. At present, however, the gluon distribution is mostly  
determined from scaling violations of $F_2$, so within the parton  
family of schemes the choice of gluon factorization is immaterial.  
Here we will fix the scheme by assuming that all moments satisfy the  
relations between the parton--scheme gluon and the \MS\ quark and  
gluons imposed by momentum conservation on second  
moments~\cite{DFLM}. This is the prescription used in common parton  
sets, and usually referred to as DIS scheme.  Explicit expressions of  
Altarelli-Parisi kernels in the DIS scheme are given in  
Appendix~\ref{schemechange}.  
  
With this prescription, the phenomenology of scaling violations can be  
studied by computing a sufficiently large  
number of truncated moments of structure  
functions, so as  to guarantee the required accuracy. If the aim is, for  
instance, a determination of $\alpha_s$ from nonsinglet scaling  
violations, all we need is a large enough number of truncated moments  
of the nonsinglet structure function.  Once an interpolation of the  
data in the measured region is available, the determination of such  
truncated moments is straightforward. This interpolation can be  
performed in an unbiased way using neural networks, as already  
mentioned in the introduction.   
One may wonder, however, whether the need to use the values  
of very high moments wouldn't be a problem. Indeed, very high moments  
depend strongly on the behavior of the structure function at large  
$y$, which is experimentally known very poorly. Furthermore, it seems  
contradictory that scaling violations of the lowest moments   
should be most dependent on the structure function  
at large $y$.  
  
This dependence is only apparent, however. Indeed, \eq{finsyst}  
for, say, the nonsinglet first moment can be rewritten as  
\beq  
{d\over dt} q_1(x_0,Q^2) = {\alpha_s(Q^2)\over2\pi} \sum_{p=0}^\infty   
{g_p^1 (x_0)  \over p!} {\hat q}_1^p(x_0,Q^2)~,  
\label{hatmom1}  
\eeq  
where  
\beq  
{\hat q}_k^p(x_0,Q^2) = \int_{x_0}^1 dy y^{k-1} (y-1)^p q(y,Q^2)~.  
\label{hatmom2}  
\eeq  
The need to include high orders in the expansion in \eq{finsyst2} is  
due to the slow convergence of the series in \eq{hatmom1}, in turn  
determined by the fact that $G_n(x_0/y)$ diverges logarithmically at  
$y = x_0$. Correspondingly, the right-hand side of the evolution  
equation depends significantly on $\hat q_1^p$ for large values of  
$p$, which signal a sensitivity to the value of $q(y,Q^2)$ in the  
neighborhood of the point $y = x_0$.  The dependence on high truncated  
moments $q_n$ is introduced when $\hat q_1^p$ is re-expressed in terms  
of $q_n$, by expanding the binomial series for $(y-1)^p$. Since this  
re-expansion is exact, it cannot introduce a dependence on the large  
$y$ region which is not there in the original expression.  The high  
orders of the expansion do instead introduce a significant dependence  
on the value of the structure function in the neighborhood of $x_0$,  
which can be kept under control provided $x_0$ is not too small,  
{\it i.e.} well into the measured region. There is therefore no obstacle  
even in practice in performing an accurate determination of $\alpha_s$  
from scaling violations of truncated moments.  
  
Let us now consider a second typical application of our method, namely  
the determination of truncated moments of the gluon distribution. In  
particular, the physically interesting case is the lowest integer  
moment, {\it i.e.} the momentum fraction in the unpolarized case or the spin  
fraction in the polarized case. The need to include a large number of  
terms in the expansion of the evolution equations seems to imply the  
need to introduce an equally large number of parameters, one for each  
gluon truncated moment. This would be problematic since it is appears  
unrealistic to fit a very large number of parameters of the gluon from  
currently available data on scaling violations.  We may, however, take  
advantage of the fact that the dependence on high order truncated  
moments is fictitious, as we have just seen, and it rather indicates  
an enhanced sensitivity to the value of $q(y)$ as $y\to x_0$.  This  
suggests that a natural set of parameters to describe the gluon  
distribution should include the first several truncated moments, as  
well as further information on the behavior of the distribution around  
the truncation point $x_0$, such as the value of the distribution (and  
possibly of some of its derivatives) at the point $x_0$.  
  
To understand how such a parametrization might work, notice that if  
$q(y)$ is regular around $y=x_0$, then it is easy to prove that  
\beq  
\lim_{p\to \infty} {\int_{x_0}^1 d y (y - 1)^p q(y) \over  
q(x_0) \int_{x_0}^1 d y (y - 1)^p } = 1~,  
\eeq  
by Taylor expanding $q(y)$ about $y=x_0$. We may therefore approximate  
the series  
which appears on the {\it r.h.s.} of \eq{hatmom1} by  
\beq  
S(x_0, n_0)  =  \sum_{p = 0}^{n_0 - 1} \frac{g_p^1 (x_0)}{p!}   
\int_{x_0}^1 d y (y - 1)^p q(y)  + \sum_{p = n_0}^\infty   
\frac{g_p^1 (x_0)}{p!} q(x_0) \int_{x_0}^1 d y (y - 1)^p. \label{appro} \\  
\eeq  
Equation~(\ref{appro}) describes the evolution of the first truncated  
moment of $q(y)$ in terms of the first $n_0$ truncated moments and of  
the value of $q(y)$ at the truncation point $x_0$.  Of course, the  
approximation gets better if $n_0$ increases. It is easy to check that  
when $x_0=0.1$ the accuracy is already better than 10\% when $n_0\sim  
7$.  This means that a parametrization of the distribution in terms of  
less than ten parameters is fully adequate. It is easy to convince  
oneself that this estimate is reliable, and essentially independent of  
the shape of the distribution $q(y)$.  In fact, because slow  
convergence arises due to the logarithmic singularity in $G_n(x_0/y)$,  
we can estimate the error of the approximation in eq.~(\ref{appro}) by  
replacing the functions $g_p^1 (x_0)/p!$ with the coefficients of the  
Taylor expansion of $\log(1 - x_0/y)$ in powers of $y - 1$, which   
we may denote by $\hat{g}_p^1 (x_0)/p!$. The error is then  
\beqn  
&&\left|\sum_{p = n_0}^\infty \frac{\hat{g}_p^1 (x_0)}{p!}  
\int_{x_0}^1 d y (y - 1)^p \left( q(y) - q(x_0) \right)  
\right|   
\label{bound} \\   
&&\phantom{aaaaaaaaaaaaaaaaa}  
\leq  
\int_{x_0}^1 d y \left| \log (1 - x_0/y) -  
\sum_{p = 0}^{n_0 - 1} \frac{\hat{g}_p^1 (x_0)}{p!} (y - 1)^p \right|  
\left| q(y) - q(x_0) \right|~.  
\nonumber  
\eeqn  
The expression inside the first absolute value on the {\it r.h.s.} of  
eq.~(\ref{bound}) is just the error made in approximating the  
logarithm with its Taylor expansion around $y = 1$; thus, it is a  
slowly decreasing function of $n_0$, it is integrable, and the  
integral receives the largest contribution from the region $y \sim  
x_0$; the second absolute value, on the other hand, is a bounded  
function of $y$ in the range $x_0 \leq y \leq 1$, which vanishes as $y  
\to x_0$ for any choice of $q(y)$. These two facts combine to limit  
the size of the error.  One can check directly that, choosing for  
example $q(y)=(1-y)^4$ as in the previous section, the accuracy is  
better than 10\% with $n_0\sim 10$ and $x_0=0.1$, in agreement with  
the previous estimate. One may also verify that, as expected, changing  
the shape of $q(y)$ does not significantly affect the result.  
  
We conclude that there is no difficulty in using the evolution of  
truncated moments for a direct extraction of the lowest truncated  
moment of the gluon distribution, provided only the higher moments of  
the distributions, which are auxiliary quantities needed in the  
extraction, are parametrized in an effective way. We have seen  
that this is possible with a reasonably small number of parameters.  
  
\sect{Outlook}   
  
In this paper, we have discussed the solution of the Altarelli-Parisi  
evolution equations of perturbative QCD by projecting parton  
distributions on a basis of truncated moments. We have seen that  
truncated moments give us a compromise between standard Mellin  
moments, which satisfy simple linear evolution equations, but are not  
measurable, and parton distributions themselves, which are measurable,  
but satisfy integro-differential evolution equations.  In this
respect, projecting on a basis of truncated moments is akin to
projecting on a basis of orthogonal polynomials~\cite{orto}, but with
the added advantage that truncated moments are physical observables.
We have further shown  
that evolution equations for truncated moments can be solved to  
arbitrarily high accuracy by using a sufficiently large basis of  
truncated moments, and we have discussed how this formalism can be  
exploited to perform a model-independent, unbiased analysis of scaling  
violations, which does not rely on parton parametrizations. We have  
also collected all technical tools which are needed for this analysis,  
and discussed the reliability of the approximations which are  
necessary in order to implement it in practice.  Such an analysis  
could be used for the determination of $\alpha_s$ and for the direct  
measurement of the contribution to the moments of the gluon  
distribution from the experimentally accessible region. In both cases,  
it would offer the advantage of allowing reliable estimates of the  
uncertainty on the result. In order to be effective, these  
phenomenological applications will need an unbiased interpolation of  
the data, such as could be achieved by means of neural  
networks. Phenomenological studies along these lines are currently  
under investigation and will be presented in forthcoming publications.

\vskip 1cm  
  
{\large {\bf Acknowledgements}}  
\noindent  
We thank Ll.~Garrido and J.I.~Latorre for discussions on neural networks  
and their applications to structure functions. We thank R.~Ball for  
a careful reading of the manuscript and several useful comments.  
  
\vskip 1.5cm  
  
\appendix  
  
\sect{Solution of matrix evolution equations at NLO}   
\label{apnlo}  
In this Appendix, we find  the general solution of the equation  
\beq  
\label{APi}  
\frac{d}{d\tau}\,q = C\,q\,,  
\eeq  
where $q$ is a vector with $M$ components, and $C$ is a generic $M \times M$  
matrix. The usual QCD evolution equations are special cases of this  
equation in which $M\le2$.  
We will assume that $C$ has a perturbative expansion in powers  
of a parameter $a(\tau)$:  
\beq  
\label{Cexp}  
C=C_0+a(\tau)C_1+\ldots~,  
\eeq  
with  
\beq  
\frac{d a(\tau)}{d\tau}=  
-b_0\,a\,\left(1+b_1\,a\,+\ldots\right)\,.  
\label{rge}  
\eeq  
For QCD applications  
\beq  
a=\frac{\as}{2\pi};\;\;\;\tau=\frac{1}{2\pi}\int_{t_0}^tdt' \as(t')~,  
\eeq  
with $t=\log(Q^2/\Lambda_{\sss QCD}^2)$, and  
\beq  
\frac{b_0}{2\pi}=\frac{33-2n_f}{12\pi}~;\quad  
\frac{b_1}{2\pi}=\frac{153-19n_f}{2\pi(33-2n_f)}~.  
\eeq  
  
The solution of eq.~(\ref{APi}) can be obtained perturbatively. Expanding  
$q$ to order $a$,  
\beq  
q=q_0+a\,q_1\;,  
\eeq  
we find  
\beqn  
\label{LO}  
&&\frac{d}{d\tau}\,q_0=C_0\,q_0~,  
\\  
\label{NLO}  
&&\frac{d}{d\tau}\,q_1 = (C_0+b_0)\,q_1+C_1 q_0\,.  
\eeqn  
The solutions of eqs.~(\ref{LO},\ref{NLO}) are  
\beqn  
&&q_0(\tau)=R^{-1}e^{\gamma\tau}R\,q_0(0)~,  
\\  
&&q_1(\tau)  
=R^{-1}e^{(\gamma+b_0)\tau}R\,q_1(0)+  
R^{-1}e^{(\gamma+b_0)\tau}\int_0^\tau d\sigma\, e^{-(\gamma+b_0)\sigma}  
{\hat C}_1 e^{\gamma\sigma}\,R\,q_0(0)\;,  
\eeqn  
where the matrix $R$ diagonalizes $C_0$,  
\beq  
RC_0R^{-1}={\rm diag}(\gamma_1,\ldots,\gamma_M)\equiv \gamma~,  
\eeq  
and   
\beq  
{\hat C}_1=RC_1R^{-1}\,.  
\eeq  
  
Collecting these results, and noting that $a\exp(b_0\tau)=a(0)$, up to  
terms of order $a^2$, we can write the solution as  
\beq  
\label{q1ter}  
q(\tau)\equiv U(C,\tau)q(0)=R^{-1}\left[e^{\gamma\tau}  
+ae^{(\gamma+b_0)\tau}\int_0^\tau d\sigma\, e^{-(\gamma+b_0)\sigma}  
{\hat C}_1 e^{\gamma\sigma}\right]R\,q(0)\,,  
\eeq  
with the initial condition  
\beq  
q(0)=q_0(0)+a(0)q_1(0)\,.  
\eeq  
The explicit expression of $U(C,\tau)$ is  
\beq  
U_{ij}(C,\tau)= R_{im}^{-1} \left[ \delta_{mn} e^{\gamma^n \tau}  
+ a(\tau)\, {\hat C}_1^{mn}\,\frac  
{e^{\gamma^n\tau}-e^{(\gamma^m+b_0)\tau}}  
{\gamma^n - \gamma^m - b_0}  \right] R_{nj}~,  
\eeq  
which, expanded to next-to-leading order reduces to  
\beqn  
&&U_{ij}(C,\tau)= R_{im}^{-1}\Bigg\{\delta_{mn}  
\left(\frac{a(0)}{a(\tau)}\right)^{\gamma_n/b_0}  
\nonumber\\  
&&\phantom{aaaaaaaaa}  
+\frac{{\hat C}_1^{mn}-b_1\gamma_n\delta_{mn}}{\gamma^m-\gamma^n+b_0}   
\left[  
a(0)\left(\frac{a(0)}{a(\tau)}\right)^{\gamma_m/b_0}  
-a(\tau)\left(\frac{a(0)}{a(\tau)}\right)^{\gamma_n/b_0}   
\right] \Bigg\}  
R_{nj}~.  
\nonumber\\  
\label{singletsol}  
\eeqn  
In the case of standard QCD evolution equations the matrix $C_0$ is at  
most $2\times2$ and is easily diagonalized. In the cases treated in  
this paper, the matrix $C_0$ is triangular and can be diagonalized  
using the methods discussed in the next Appendix.  
  
\sect{Diagonalization of triangular matrices}  
\label{triangular}  
In this Appendix, we show how to construct the matrix $R$ which  
diagonalizes a generic $n\times n$ triangular matrix $T$ by means of  
the recursion relations eqs.~(\ref{trir},\ref{invtrir}). The matrix  
$R$ is defined by the requirement that  
\beq   
R T R^{-1}=\mbox{diag}(\gamma_1,\dots,\gamma_n)~,  
\label{tridiag}  
\eeq  
where the matrix $T$ is upper triangular, {\it i.e.} $T_{ij}=0$ if $i>j$.  
It is easy to see, by solving the secular equation, that the eigenvalues  
$\gamma_i$ of $T$ coincide with its diagonal elements,   
\beq  
\gamma_i=T_{ii}~.  
\label{evals}  
\eeq  
Now, define eigenvectors $v^j$ associated to the $j$-th eigenvalue  
$T_{jj}$, with components $v_i{}^j$:   
\beq  
\sum_{k=1}^n T_{i k} v_k{}^j=\gamma_j v_i{}^j~.  
\label{eveccon1}  
\eeq  
Clearly, the matrix $R^{-1}$ coincides with the matrix of right  
eigenvectors, $(R^{-1})_{ij}= v_i{}^j$, while the matrix $R$ coincides  
with the matrix of left eigenvectors $\sum_{k=1}^n \hat v^j{}_k T_{k  
i} =\gamma_j \hat v^j{}_i$, $R_{ij}={\hat v}^i{}_j$.  The eigenvector  
condition eq.~(\ref{eveccon1}) immediately implies that the $j$-th  
component of the $j$-th eigenvector is equal to one:  
$v_j{}^j=1$. Furthermore, it is clear that eq.~(\ref{eveccon1}) can  
only be satisfied if all components $v_k{}^j$ of the $j$-th  
eigenvector with $k>j$ vanish,  
\beq  
v_j{}^j=1~; \qquad v_k{}^j=0 \quad \mbox{if} \quad k>j~.  
\label{evecs}  
\eeq  
Using eq.~(\ref{evecs}) and the fact that the matrix $T$ is  
triangular, eq.~(\ref{eveccon1}) can be written as  
\beq  
\sum_{k=i}^j T_{i k} v_k{}^j=\gamma_j v_i{}^j~.  
\label{eveccon}  
\eeq  
Substituting the explicit form of the eigenvalues, eq.~(\ref{evals}),  
and identifying $v_i{}^j =(R^{-1})_{ij}$, this is immediately seen to  
coincide with eq.~(\ref{invtrir}). Furthermore, using the condition  
$v_j{}^j=1$, this equation can be viewed as a recursion relation which  
allows the determination of the $(k-1)$-th element of $v^j$ once the  
$k$-th element is known, which is what we set out to prove. The same  
argument, applied to the left eigenvectors, leads to the expression in  
eq.~(\ref{trir}) for $R$.  
  
\sect{Splitting Functions in the DIS scheme}  
\label{schemechange}  
\def\l{\left}  
\def\r{\right}  
\def\ione{one}   
\def\itwo{\log x}  
\def\ithree{\log^2 x}  
\def\ifour{\log(1 - x)}  
\def\ifive{\log x \log(1 - x)}  
\def\isix{\frac{\log x}{1 - x}}  
\def\iseven{\frac{\log^2 x}{1 - x}}  
\def\ieight{\left(\frac{1}{1-x}\right)_+}  
\def\inine{\frac{\log x \log(1 - x)}{1 - x}}  
\def\ionefive{\left(\frac{\log (1-x)}{1-x}\right)_+}  
\def\ioneseven{\log^2(1 - x)}  
\def\ioneeight{{\rm Li}_2 (x)}  
\def\ionenine{\left(\frac{\log^2 (1-x)}{1-x}\right)_+}  
\def\nf{n_f}  
\def\ca{C_A}  
\def\cf{C_F}  
In this Appendix, we give the explicit expressions of the  
Altarelli-Parisi kernels in the DIS scheme~\cite{aem,DFLM}. We define the  
non-singlet splitting functions as  
\beq  
P_\pm (x) \equiv P_{qq} (x) \pm P_{q{\bar q}}(x) =  
P_\pm^{(0)}(x) + a P_\pm^{(1)}(x) + \ldots  
\eeq  
and the singlet $2\times 2$ matrix of splitting functions as  
\beq  
P(x)=P^{(0)}(x)+aP^{(1)}(x)+\ldots\,.  
\eeq  
The \MS\ LO and NLO kernels are given for example in eqs.~(4.94) and  
(4.102)-(4.112) of ref.~\cite{kis}, whose notation and conventions we  
have followed throughout this paper. The splitting functions in the  
DIS scheme can be constructed from these by a change of factorization  
scheme.  
  
To next-to-leading order, a generic change of factorization scheme for  
the splitting functions is  
\beqn  
&& P_{\pm}^{(1)} \rightarrow P_{\pm}^{(1)} - b_0  E^{\sss NS}~,  
\label{scns}  
\\  
&& P^{(1)} \rightarrow P^{(1)} + [E,P^{(0)}] - b_0 E\,,  
\label{scs}  
\eeqn  
where $E$ and $E^{\sss NS}$ are functions of $x$, and the commutator  
in eq.~(\ref{scs}) is defined in terms of convolutions, as  
\beq  
\left[E,P^{(0)}\right]=E\otimes P^{(0)}-P^{(0)}\otimes E\,,  
\eeq  
and is thus in general nontrivial to compute.  
  
The transformation that takes from the $\overline{\mathrm{MS}}$ to the
DIS scheme, defined as in ref.~\cite{DFLM}, is given by \beq E^{\sss
NS}(x)=C^q(x)\,, \eeq \beqn E(x)=\left[ \matrix{ C^q(x) & 2 n_f C^g(x)
\cr - C^q(x) & -2 n_f C^g(x)}\right]\,, \eeqn where $C^q(x)$ and
$C^g(x)$ are the next-to-leading terms of the quark and gluon
coefficient functions for the unpolarized deep-inelastic structure
function $F_2(x,Q^2)$ in the $\overline{\rm MS}$ scheme:
\beqn
F_2(x,Q^2)&=&
x\,\langle e^2\rangle\int_x^1 \frac{dy}{y}\left\{
\left[\delta\left(1-\frac{x}{y}\right)+a C^q\left(\frac{x}{y}\right)
\right]\,\Sigma(y,Q^2)
+2\nf a C^g\left(\frac{x}{y}\right)g(y,Q^2)\right\}
\nonumber\\ 
&+&x\,\int_x^1
\frac{dy}{y} \left[\delta\left(1-\frac{x}{y}\right)
+a C^q\left(\frac{x}{y}\right) \right]\,q_{\sss NS}(y,Q^2)+\ldots~,
\eeqn
where $\langle e^2 \rangle=\sum_{i=q,{\bar
q}}e^2_i/(2\nf)$.
Explicitly,
\beqn
C^q(x)&=&
\cf\Bigg[2\ionefive-\frac{3}{2}\ieight-(1+x)\log(1-x)
\nonumber\\&&\phantom{aaa} -\frac{1+x^2}{1-x}\log x+3+2x
-\left(\frac{\pi^2}{3}+\frac{9}{2}\right)\delta(1-x) \Bigg]~, \\
C^g(x)&=&
T_R\Bigg[\left((1-x^2)+x^2\right)\log\frac{1-x}{x}-8x^2+8x-1\Bigg]\,.
\eeqn 
Notice that the expression for the scheme change given in eq.~(3.9)
of ref.\cite{DFLM} lacks a factor of $2\nf$ in the $qg$ entry because
it refers to a single flavor.

The explicit expression of the commutator is  
\begin{eqnarray}  
\left[E,P^{(0)}\right]_{qq}&=&  
\cf\,\nf\,
\Bigg[ -3\,\ioneeight
-\frac{1}{2}\,(1 + 4\,x - 4\,x^2)\,\left(\ithree-\frac{\pi^2}{3}\right)  
\nonumber \\  
& + & 
\Bigg(\frac{4}{3\,x}-\frac{5}{2}+7\,x-\frac{22\,x^2}{3}\Bigg)\, \ifour  
- \Bigg(\frac{3}{2}+9\,x - \frac{22\,x^2}{3} \Bigg)\,\itwo  
\nonumber\\
& + &  ( 1 - 2\,x + 2\,x^2 )\,\log(1-x)\log\frac{1-x}{x^2}
 + \frac{2}{3\,x}  
    - \frac{55}{6} 
    + \frac{20\,x}{3} 
    - \frac{8\,x^2}{3}
 \Bigg]~~,  
\label{firstc}  
\end{eqnarray}  
\begin{eqnarray}  
\left[E,P^{(0)}\right]_{qg}  & = & 
2\,\nf\,b_0\,C^g(x)  
+ {\nf}^2\,\Bigg[  
           ( 1 + 4\,x + 4\,x^2 )\,
           \left(\ioneeight+\frac{1}{2} \ithree-\frac{\pi^2}{6}\right)
\nonumber \\
           & - & ( 2 + 4\,x - 6\,x^2 )\,\ifour  
           +  ( 2 + 16\,x + 10\,x^2 )\,\itwo  
\nonumber \\
           & + & \frac{11}{2} + 24\,x  - \frac{59\,x^2}{2}  \Bigg]  
\nonumber \\  
& + & \cf\,\nf\, \Bigg[ -(1 - 2\,x+2\,x^2)\,
                 \left(2\,\ioneeight+\ioneseven-\frac{\pi^2}{3}\right)
\nonumber \\   
               & - & ( 1 + 10\,x - 10\,x^2 )\,\ifour  
               -  (1 - 4\,x + 10\,x^2 )\,\itwo  
               - 6 + 11\,x   - 8x^2 \Bigg]  
\nonumber \\   
& + & 2\,\ca\,\nf\,\Bigg[ -(1 + 4x )\,
                    \left(\ioneeight+\frac{1}{2}  \,\ithree\right)
\nonumber \\
                    & + & \Bigg(\frac{2}{3x} -\frac{1}{2}  
                    + 12x - \frac{79x^2}{6} \Bigg)\,\ifour  
\nonumber \\   
                    & + & ( 1 - 2x + 2x^2)\,\log(1-x)\log\frac{1-x}{x}  
                    - \Bigg(\frac{1}{2} + 16x - \frac{31x^2}{6} \Bigg)\,\itwo  
\nonumber \\
                    & + & \frac{1}{3x}-\frac{43}{12}   - \frac{121x}{6}
                    + \pi^2 x  + \frac{281 x^2}{12} - \frac{\pi^2 x^2}{3}  
                    \Bigg]  
\nonumber \\   
\label{secc}  
\end{eqnarray}  
\begin{eqnarray}  
\left[E,P^{(0)}\right]_{gq} & = & b_0\,C^q(x) + \cf\,\nf\,  
\Bigg[ ( 1 + x )\,(\ithree  + 2\, \ioneeight)  
\nonumber \\  
& + & \frac{1}{3}\Bigg(20 - \pi^2 - \frac{2}{x} - 2x - \pi^2 x - 16x^2\Bigg)  
\nonumber \\   
& + & \Bigg( 1 + 5x - \frac{4x^2}{3} \Bigg)\,\itwo  
- \Bigg( 1 + \frac{4}{3x} - x -  \frac{4x^2}{3} \Bigg)\,\ifour  \Bigg]  
\nonumber \\   
& + & \cf^2\,\Bigg[ \Bigg(\frac{27}{4}-  4\zeta(3)\Bigg) \,\delta(1-x)  
+ 3\,\ionefive + \frac{3}{2}\,\ithree  
\nonumber \\   
& - & \Bigg( 12 - \frac{6}{x} + \frac{5x}{2} \Bigg)\,\ifour  
+ \Bigg( 2 - \frac{2}{x} - x \Bigg)\,\ioneseven  
- \frac{9}{2}-\frac{\pi^2}{2} + \frac{3x}{2}   
\nonumber \\   
& - & 2 \Bigg( 2 - \frac{2}{x} - x \Bigg)\,\ifive  
- \Bigg( 1 - \frac{4}{x} - 2x \Bigg)\,\ioneeight  
+ \Bigg( \frac{7}{2} + 4\,x \Bigg)\,\itwo  
\nonumber \\   
& - & ( 1 + x^2)\left[3\,\ionenine  + \iseven - 4\,\inine \right]\,  
\nonumber \\   
& - & \frac{3}{2}(1-x^2)\,\isix  
+ \Bigg( \frac{27}{4} + \frac{2\,{\pi }^2}{3} +  
\frac{9x^2}{2} + \frac{2\pi^2x^2}{3} \Bigg)\,\ieight\Bigg]  
\nonumber \\   
& + & \ca\,\cf\,\Bigg[ -6\,\ionefive + 3\,\isix  
+ \Bigg(\frac{\pi^2}{2}+ 4\zeta(3)\Bigg)  \,\delta(1-x)  
\nonumber \\   
& + & 2\Bigg( 1 - \frac{2}{x} - 2\,x \Bigg)\,\ioneeight  
- 8 x \,\inine + 6x\,\ionenine  
\nonumber \\   
& + & 2 x \,\iseven -\Bigg( 9x + \frac{4\pi^2x}{3} \Bigg)\,\ieight  
- (1 + 2 x^2)(3\,\itwo+\ithree)  
\nonumber \\   
& + & \Bigg( \frac{2}{x} - 2\,x^2 \Bigg)\,\ioneseven  
+ \Bigg( 2 - \frac{4}{x} - 2\,x + 4\,x^2 \Bigg)\, \ifive  
\nonumber \\   
& + & \Bigg( 17 - \frac{6}{x} - 4x + 6x^2 \Bigg)\,\ifour  
- 2 - 6x + \frac{\pi^2x}{3} + 8x^2 + \frac{2\pi^2x^2}{3}  \Bigg]~.  
\label{thirdc}  
\end{eqnarray}  
Finally  
\beq  
\left[E,P^{(0)}\right]_{gg} = - \left[E,P^{(0)}\right]_{qq}~~.  
\label{lastc}  
\eeq  
Note that $\left[E,P^{(0)}\right]_{qg}$ denotes the $qg$ matrix element
of the commutator; i.e., using
the conventions of ref.~\cite{kis}, upon scheme change 
\beq
2\nf P^{(1)}_{qg}\to 2\nf P^{(1)}_{qg}+\left[E,P^{(0)}\right]_{qg}- b_0 E_{qg}~.
\eeq

\sect{Truncated moment integrals}  
\label{integrals}  
\def\ione{}   
\def\itwo#1{\log #1}  
\def\ithree#1{\log^2 #1}  
\def\ifour#1{\log(1 - #1)}  
\def\ifive#1{\log #1 \log(1 - #1)}  
\def\isix#1{\frac{\log #1}{1 - #1}}  
\def\iseven#1{\frac{\log^2 #1}{1 - #1}}  
\def\ieight#1{\left(\frac{1}{1-#1}\right)_+}  
\def\inine#1{\frac{\log #1 \log(1 - #1)}{1 - #1}}  
\def\iten#1{\frac{1}{1+#1}}  
\def\ioneone#1{\frac{\log #1 \log (1+#1)}{1+#1}}  
\def\ionetwo#1{\frac{\log^2 #1}{1+#1}}  
\def\ionethree#1{\frac{{\rm Li}_2 (-#1)}{1+#1}}  
\def\ionefour#1{\log #1 \log (1+#1)}  
\def\ionefive#1{\left(\frac{\log (1-#1)}{1-#1}\right)_+}  
\def\ionesix#1{{\rm Li}_2 (-#1)}  
\def\ioneseven#1{\log^2(1 - #1)}  
\def\ioneeight#1{{\rm Li}_2 (#1)}  
\def\ionenine#1{\left(\frac{\log^2 (1-#1)}{1-#1}\right)_+}  
  
The integrals which are needed in order to compute the Mellin moments  
of the NLO Altarelli-Parisi splitting functions are  
well known~\cite{devduke}. Here we list all the truncated moment  
integrals which are necessary in order to determine the evolution  
kernels $G_n(x)$, eq.~(\ref{kern}), from the expressions of the  
splitting functions given in appendix~C. The triangular anomalous  
dimension matrices $C_{kl}$, eq.~(\ref{matr0}), can be easily determined  
from these formulas by Taylor expansion.  
\beqn  
\int_x^1dz\, z^{n-1} &=& \frac{1 - x^n}{n}~,  
\label{i1}  
\\  
\int_x^1dz\, z^{n-1} \itwo{z} &=& - \frac{1 - x^n\,(1 - n\,\log x)}{n^2}~,  
\label{i2}  
\\  
\int_x^1dz\, z^{n-1} \ithree{z} &=& \frac{2 - x^n\,( 2 - 2\,n\,\log x +   
    n^2\,\log^2 x ) }{n^3}~,  
\label{i3}  
\\  
\int_x^1dz\, z^{n-1} \ifour{z} &=& -\frac{1}{n\,(n+1) }  
    \Bigg[ x^{n+1}\,{_2 F_1}(n+1,1;n+2;x)   
\nonumber \\&&   
+  (n+1) \,\Big( \gamma_E + x^n\,\log (1 - x)   
+ \psi^{(0)}(n+1) \Big)  \Bigg]~,  
\label{i4}  
\\  
\int_x^1dz\, z^{n-1} \ifive{z} &=&   
       -\frac{1}{n^2}\Bigg[-\gamma_E-n\,x^{n+1}\,  
       \Phi(x,2,n+1) - x^n\,\log (1 - x)   
\label{i5} \\&&   
       +n\,x^n\,\log (1 - x)\,\log x + x^{n+1}\,\Phi(x,1,n+1)\,(n\,\log x-1)    
\nonumber \\&&   
+ \psi^{(0)}(n) + n\,[\psi^{(0)}(n)]^2 - n\,[\psi^{(0)}(n+1)]^2 +   
       n\,\psi^{(1)}(n+1) \Bigg]~,  
\nonumber  
\\  
\int_x^1dz\, z^{n-1} \isix{z} &=& \frac{x^n}{n^2} + x^{n+1}\,  
    \Phi(x,2,n+1) - \frac{x^n}{n}\,\log x   
\nonumber \\&&   
-   x^{n+1}\,\Phi(x,1,n+1)\,\log x -   
   \psi^{(1)}(n)~,  
\label{i6}  
\\  
\int_x^1dz\, z^{n-1} \iseven{z} &=& -\frac{2\,x^n}{n^3} - 2\,x^{n+1}\,  
   \Phi(x,3,n+1) + \frac{2\,x^n}{n^2}\,\log x   
\nonumber \\&&   
    +2\,x^{n+1}\,\Phi(x,2,n+1)\,\log x -\frac{x^n}{n} \,\log^2 x  
\nonumber \\&&   
    -x^{n+1}\,\Phi(x,1,n+1)\,\log^2 x -   
   \psi^{(2)}(n)~,  
\label{i7}  
\\  
\int_x^1dz\, z^{n-1} \ieight{z} &=& -\gamma_E - \frac{x^n}{n}\,  
      {_2 F_1}(1,n;n+1;x) -   
   \psi^{(0)}(n)~,  
\label{i8}  
\\  
\int_x^1dz\, z^{n-1} \inine{z} &=& {\rm Li}_3 (1 - x) -   
       \log (1 - x) {\rm Li}_2 (1 - x)   
\nonumber \\&&   
- \sum_{k=1}^{n-1} \int_x^1 dz\, z^{k-1} \ifive{z}~,  
\label{i9}  
\\  
\int_x^1dz\, z^{n-1} \iten{z} &=& - \frac{x^n}{n}\,{_2 F_1}(n,1;n+1;-x)  
\nonumber \\&&   
       - \frac{1}{2} \left[\psi^{(0)}\left(\frac{n}{2}\right) -   
         \psi^{(0)}\left(\frac{n+1}{2}\right)\right]~,  
\label{i10}  
\\  
\int_x^1dz\, z^{n-1} \ioneone{z} &=& (-1)^{n-1} \Bigg[- \frac{\zeta (3)}{8} -   
        \frac{\log x \log^2 (1 + x)}{2} + \frac{\log^3(1 + x)}{3}   
\nonumber \\&&   
+ \log(1 + x) {\rm Li}_2 \left(\frac{x}{1 + x}\right) +   
        {\rm Li}_3 (-x) + {\rm Li}_3 \left(\frac{x}{1 + x}\right)   
\nonumber \\&&   
        -\sum_{k=1}^{n-1} (-1)^{k-1}\int_x^1dz\, z^{k-1} \ionefour{z} \Bigg]~,  
\label{i11}  
\\  
\int_x^1dz\, z^{n-1} \ionetwo{z} &=&   
   -\frac{2\,x^n}{n^3} + 2\,x^{n+1}\,\Phi(-x,3,n+1)  
\nonumber \\&&   
   +\frac{2}{n^2}\,x^n\,\log x - 2\,x^{n+1}\,\Phi(-x,2,n+1)\,\log x  
\nonumber \\&&   
   -\frac{x^n}{n} \,\log^2 x + x^{n+1}\,\Phi(-x,1,n+1)\,\log^2 x   
\nonumber \\&&   
-   \frac{1}{8}\left[\psi^{(2)}\left(\frac{n}{2}\right) -   
   \psi^{(2)}\left(\frac{n+1}{2}\right)\right]~,  
\label{i12}  
\\  
\int_x^1dz\, z^{n-1} \ionethree{z}   
&=& (-1)^{n-1} \Bigg[- \frac{7 \zeta (3)}{4}-    
        \frac{\pi^2}{12} \log2  
+    \log (1 + x) {\rm Li}_2 (-x)   
\nonumber \\&&   
         + \log x \log^2(1 + x) + \frac{\pi^2}{3} \log (1 + x)   
         + 2 {\rm Li}_3 \left(\frac{1}{1 + x}\right)   
\nonumber \\&&   
- \frac{\log^3(1 + x)}{3} -  
        \sum_{k=1}^{n-1} (-1)^{k-1}\int_x^1dz\, z^{k-1} \ionesix{z} \Bigg]~,  
\label{i13}  
\\  
\int_x^1dz\, z^{n-1} \ionefour{z} &=& \frac{1}{4\,n^2}\Bigg[-4\,n\,x^{n+1}\,  
     \Phi(-x,2,n+1) - 4\,\log 2   
\nonumber \\&&  
+  4\,x^{n+1}\,\Phi(-x,1,n+1)\,  
      ( -1 + n\,\log x )  + 4\,x^n\,\log (1 + x)   
\nonumber \\&&  
-   4\,n\,x^n\,\log x\,\log (1 + x)   
+     2\,\psi^{(0)}\left(1 + \frac{n}{2}\right)   
-   2\,\psi^{(0)}\left(\frac{n+1}{2}\right)   
\nonumber \\&&  
-  n\,\psi^{(1)}\left(1 + \frac{n}{2}\right)   
+  n\,\psi^{(1)}\left(\frac{n+1}{2}\right)\Bigg]~,  
\label{i14}  
\\  
\int_x^1dz\, z^{n-1} \ionefive{z} &=&    
( n-1) \, ( 1 - x) \,\Big[{_4 F_3}(1,1,1,2 - n;2,2,2;1-x)   
\label{i15} \\&& -{_3 F_2}(1,1,2-n;2,2;1-x)\,  
       \log (1 - x) \Big] +\frac{{\log (1 - x)}^2}{2}~,  
\nonumber  
\\  
\int_x^1dz\, z^{n-1} \ionesix{z} &=& \frac{1}{n^2\,(n + 1) }  
   \Bigg[x^{n + 1}\,{_2 F_1}(n+1,1;n+2;-x)   
\nonumber \\&&  
    -(n+1) \,\Bigg( \frac{n\,{\pi }^2}{12} - \log 2 +   
        x^n\,\log (1 + x)   
\label{i16} \\&&  
+ \frac{1}{2}  
\left[\psi^{(0)}\left(1 +\frac{n}{2}\right)  
     -\psi^{(0)}\left(\frac{n+1}{2}\right)\right] +   
        n\,x^n\,{\rm Li}_2 (-x) \Bigg)\Bigg]~,  
\nonumber  
\\  
\int_x^1dz\, z^{n-1} \ioneseven{z} &=&   
          2\,(1 - x) \Big[{_4F_3}(1,1,1,1-n;2,2,2;1-x)   
\nonumber \\&&  
            -{_3F_2}(1,1,1-n;2,2;,1-x)\log (1 - x)\Big]   
\nonumber \\&&  
 + \frac{1-x^n}{n}\,\log^2(1-x)~,  
\label{i17}  
\\  
\int_x^1dz\, z^{n-1} \ioneeight{z} &=& \frac{1}{n^2\,\Gamma(n+2)}  
       \Bigg[-\Gamma(2 + n)\,(\gamma_E+\psi^{(0)}(n+1))   
\nonumber \\&&  
      + \Gamma(n + 1)\,\Bigg(-x^{n+1}\,{_2 F_1}(n+1,1;n+2;x)   
\nonumber \\&&  
      +(n+1)\,\Bigg( n\,\frac{\pi^2}{6}-x^n\,\log(1 - x)   
      - n\,x^n\,{\rm Li}_2 (x) \Bigg)  
        \Bigg) \Bigg]~,  
\label{i18}  
\\  
\int_x^1dz\, z^{n-1} \ionenine{z} &=& \frac{{\log (1 - x)}^3}{3}   
   -(1-x)\sum_{j = 0}^{n - 2}  
     \Bigg[ 2\,{_4 F_3}(1,1,1,-j;2,2,2;1-x)   
\nonumber \\&&  
     - 2\,{_3 F_2}(1,1,-j;2,2;1-x)\,\log (1 - x)   
\nonumber \\&&  
+   \frac{1 - x^{j+1}}{(j+1 )\,(1-x)}\,\log^2 (1 - x)  
           \Bigg ]~,  
\label{i19}  
\eeqn  
where   
\beqn  
&&{\rm Li}_m(z)=\sum_{k=1}^\infty \frac{z^k}{k^m}~;  
\quad {\rm Li}_2(z)=\int_z^0dt\,\frac{\log(1-t)}{t}~;  
\label{polylogs}  
\\  
&&\psi^{(k)}(n)=\frac{d^{k+1}\log\Gamma(n)}{dn^{k+1}}~;  
\quad \gamma_E = -\psi^{(0)}(1) \approx 0.577216~;  
\label{polygamma}  
\\  
&&\Phi(z,s,a)=\sum_{k=0}^\infty \frac{z^k}{(a+k)^s}~;  
\label{phi}  
\\  
&&{_p F_q}(a_1,\ldots,a_p;b_1,\cdots,b_q;z)=\sum_{k=0}^\infty  
\frac{(k+a_1)\ldots(k+a_p)}{(k+b_1)\cdots(k+b_q)}\frac{z^k}{k!}\,.  
\label{hyper}  
\eeqn  
Note that the integrals in eqs.~(\ref{i9}),~(\ref{i11}),~(\ref{i13})  
and~(\ref{i19}) are valid only when $n$ is a positive integer, which  
is what is usually needed.  If an expression for real or complex $n$  
is necessary, one should compute these integrals numerically.  All  
other integrals are given in a form that immediately generalizes to  
complex $n$. The special functions in  
eqs.~(\ref{polylogs}-\ref{hyper}) are available in algebraic  
manipulation programs of common use.

\vfill\eject


\begin{thebibliography}{99}  
  
\baselineskip14pt  
  
\bibitem{qcd} See {\it e.g.}  S.~Catani {\it et al.}, {\tt hep-ph/0005025}, to be  
  published in the proceedings of the workshop ``Standard Model  
  Physics (and more) at the LHC'' (CERN, 1999).  
  
\bibitem{pdfrev} See {\it e.g.} S.~Forte, {\it Nucl. Phys.} {\bf A666} (2000)  
113.  
  
\bibitem{kis} See  {\it e.g.}  R.K.~Ellis, W.J.~Stirling and  
B.R.~Webber, ``QCD and Collider Physics'' (C.U.P., Cambridge, 1996).  
  
\bibitem{verm} S.~A.~Larin {\it et al.}, \np{B492}{97}{338}, {\tt hep-ph/9605317}.  
  
\bibitem{willy} E.~B.~Zijlstra and W.~L.~van~Neerven, \np{B383}{92}{525}.   
  
\bibitem{AP} G.~Altarelli and G.~Parisi, \np{B126}{77}{298}.  
  
\bibitem{aem} G.~Altarelli, R.~K.~Ellis and G.~Martinelli,   
{\it  Nucl. Phys.} {\bf B143} (1978) 521; {\bf B157} (1979) 461.  
  
\bibitem{cata} S.~Catani, {\it Z.~Phys.} {\bf C75} (1997) 665,   
{\tt hep-ph/9609263}.  
  
\bibitem{orto} F.J.~Yndurain, {\it Phys. Lett.} {\bf B74} (1978) 68;  
G.~Parisi, N.~Sourlas, {\it Nucl. Phys.} {\bf B151} (1979) 421;  
W.~Furmanski, R.~Petronzio, {\it Nucl. Phys.} {\bf B195} (1982) 237.  
  
\bibitem{pdfer} W.~T.~Giele and S.~Keller, \pr{D58}{98}{094023},  
{\tt hep-ph/9803393};  
D.~Kosower, W.~T.~Giele and S.~Keller, Proc. of the 1999 ``Rencontres de  
Physique de la Vall\'ee d'Aoste'', pag.~255 (INFN, Frascati, 1999);  
R.~D.~Ball and J.~Huston, in S.~Catani {\it et al.}, {\tt hep-ph/0005114}.  
  
\bibitem{abfr} G.~Altarelli {\it et al.}, {\em Nucl. Phys.} {\bf B496}, 337  
(1997), {\tt hep-ph/9701289}; {\em Acta Phys. Pol.} {\bf B29} (1998) 1145,  
{\tt hep-ph/9803237}.  
  
\bibitem{FM} S.~Forte and L.~Magnea, \pl{B448}{99}{295}, {\tt hep-ph/9812479};  
in Proceedings of the International Europhysics Conference on High-Energy   
Physics (EPS-HEP 99), Tampere 1999, {\tt hep-ph/9910421}.  
  
\bibitem{Lebe} See {\it e.g.} Encyclopedic Dictionary of Mathematics,  
ed. by S.~Iyanaga and Y.~Kawada, M.I.T. Press (Cambridge, Mass., 1980).  
  
\bibitem{DFLM} M.~Diemoz, S.~Ferroni, M.~Longo and G.~Martinelli,   
\zp{C39}{88}{20}.  
  
\bibitem{devduke} For comprehensive tables see {\it e.g.}  
A.~Devoto and D.~W.~Duke, {\it Riv. Nuovo Cim.} {\bf 7}~(1984)~1;  
J.~Bl\"umlein and S.~Kurth, {\it Phys. Rev.} {\bf D60} (1999) 014018,  
{\tt hep-ph/9810241}.  
  
\end{thebibliography}
\end{document}